\documentclass[reprint,amsmath,amssymb,aps,prx,nofootinbib,superscriptaddress,oneside]{revtex4-2}

\usepackage{graphicx}
\usepackage{color}
\usepackage{epstopdf}
\usepackage[usenames,dvipsnames]{xcolor}
\usepackage{bm}
\usepackage{paralist}
\usepackage{amsmath}
\usepackage{amsthm}
\usepackage{amssymb}
\usepackage{pstool}
\usepackage[percent]{overpic}
\usepackage{rotating}
\usepackage{lipsum}
\usepackage[normalem]{ulem}
\usepackage{titlesec}
\usepackage{fouriernc}
\usepackage[T1]{fontenc}
\usepackage{enumitem}
\usepackage{mathtools}
\usepackage{nccmath}
\usepackage{array}

\usepackage[colorlinks=true,linkcolor=Magenta,citecolor=Magenta,breaklinks=true]{hyperref}
\usepackage{orcidlink}

\setlist[itemize]{noitemsep, nolistsep}

\definecolor{Xred}{HTML}{B31B1B}

\usepackage{tikz}
\usepackage{pgfplots}

\definecolor{verylightgray}{HTML}{F3F3F3}
\definecolor{notsoverylightgray}{HTML}{E3E3E3}

\newtheorem{theorem}{Theorem}
\newtheorem*{theorem*}{Theorem}
\newtheorem*{corollary*}{Corollary}
\newtheorem*{lemma*}{Lemma}

\newtheorem*{assumption*}{Assumption}

\newtheorem*{principle*}{Principle}

\newcommand{\braket}[2]{\langle #1|#2\rangle}
\DeclarePairedDelimiter{\bra}{\langle}{\rvert}%
\DeclarePairedDelimiter{\ket}{\lvert}{\rangle}%
\DeclarePairedDelimiterX\ketbra[2]{\lvert}{\rvert}{#1\delimsize\rangle\!\delimsize\langle#2}%
\DeclarePairedDelimiterX\projector[1]{\lvert}{\rvert}{#1\delimsize\rangle\!\delimsize\langle#1}%
\DeclareMathOperator{\tr}{tr}

\begin{document}

\title{Convex conservation of von Neumann entropy implies unitarity or antiunitarity}

\author{Nicolas G. Underwood\,\orcidlink{0000-0003-4803-2629}}
\email{nick.underwood@newcastle.ac.uk}
\affiliation{Quantum Group, School of Computing, Newcastle University, 1 Science Square, Newcastle upon Tyne, NE4 5TG, UK}

\author{Jonte R. Hance\,\orcidlink{0000-0001-8587-7618}}
\email{jonte.hance@newcastle.ac.uk}
\affiliation{Quantum Group, School of Computing, Newcastle University, 1 Science Square, Newcastle upon Tyne, NE4 5TG, UK}

\begin{abstract}
\hspace{-\parindent}%
On the space of density matrices $D(H)$, Kadison's theorem establishes that all \emph{invertible} convex transformations $K:D(H)\to D(H)$ are unitarity or antiunitary.
For a finite Hilbert space, we propose a similar theorem which replaces the invertibility requirement with that of entropy conservation. 
Thus, we argue, the supposition of reversibility in quantum mechanics may be replaced with a principle of entropy conservation. 
\end{abstract}

\maketitle

\titleformat{\subsection}[runin]{\normalfont\itshape\bfseries}{{\color{Magenta}\thesection\thesubsection}}{.5em}{}[.---]

\titleformat{\section}
  {\normalfont\large\bfseries}
  {{\color{Magenta}\thesection}}
  {.5em}
  {\MakeUppercase}
  [\vspace{.5ex}\titlerule]

\titlespacing*{\subsubsection}{0ex}{1ex}{1ex}

\section{Introduction}
Unitary and antiunitary transformations are the dual conclusions of a number of related quantum symmetry theorems--see Chp.~5 of Ref.~\cite{landsmanFoundationsQuantumTheory2017} for a list and comparison of six of these.
The best known are Wigner's theorem \cite{wignerGruppentheorieUndIhre1931,bargmannNoteWignersTheorem1964} and Kadison's theorem \cite{kadisonTransformationsStatesOperator1965,hunzikerSYMMETRYOPERATIONSQUANTUM1972} (see below). 
The purpose of this short paper is to propose an information-theoretic addition to this list, albeit at present only for finite Hilbert spaces, where von Neumann entropy is traditionally defined.
(The maximally mixed state, which on finite Hilbert spaces corresponds to maximum entropy $\log N$, is not well defined on infinite-dimensional Hilbert spaces.)
This is part of a broader body of work investigating the sufficiency of entropy conservation to establish the reversibility of physical laws, with previous works discussing Gibbs/Shannon entropy for discrete classical systems \cite{underwoodSignaturesRelicQuantum2019} and differential/Jaynes entropy for continuous classical systems \cite{underwoodPrincipleInformationConservation2020}.
As entropy is widely understood as a scalar measure of information \cite{shannonMathematicalTheoryCommunication1949}, through our below theorem its conservation provides a \emph{physical} justification for unitary transformations, in line with the it-from-bit worldview.
Through Stone's theorem, this may then be extended to Schr\"{o}dinger evolution \cite{stoneLinearTransformationsHilbert1930,simonQuantumDynamicsAutomorphism1976}, so that the Schr\"{o}dinger equation follows from this conservation, at least within the context we have so far proved it.
An immediate consequence is that no-go theorems that are traditionally justified through the need to break unitarity (e.g., no-cloning \cite{woottersSingleQuantumCannot1982,dieksCommunicationEPRDevices1982}, no deletion \cite{kumarpatiImpossibilityDeletingUnknown2000}, and extensions thereof) may instead be justified on the grounds of creating or destroying information.
This is not a new observation. 
Indeed, this work could be considered an extension of Ref.~\cite{horodeckiCommonOriginNoCloning2005}, which proved this implication without arriving at a conclusion of (anti)unitarity.
We also speculate that the theorem, or an adaptation of it, may also be used to help inform recent information-theoretic reconstructions of quantum mechanics and/or generalised probabilistic theories (GPTs) in which reversibility has been axiomatised. See for instance Axiom 5 of Ref.~\cite{hardyQuantumTheoryFive2001}, the purification postulate of Ref.~\cite{chiribellaInformationalDerivationQuantum2011} (or Principle 6 in Ref.~\cite{chiribellaQuantumTheoryNamely2012}), and requirement 4 of Ref.~\cite{masanesDerivationQuantumTheory2011}. 
Finally, we speculate that in relativistic or gravitational scenarios in which unitarity may potentially be broken, it may be possible to guide its replacement through the construction of a physically reasonable information metric. 

To help compare our result with the best known unitary symmetry theorems, let $ L( H)$ be the vector space of linear operators ($N\times N$ complex matrices) on $N$-dimensional Hilbert space $ H$ with Frobenius/Hilbert-Schmidt inner product,
\begin{align}
(\rho,\sigma)_\text{HS}:=\tr(\rho^\dagger\sigma).
\end{align}
The convex set of density matrices $ D( H)$ is the subset of $ L(H)$ for which
\begin{align}\label{density_matrix_conditions}
\rho^\dagger=\rho,\quad
\tr\rho=1,\quad
(\ketbra{\psi}{\psi},\rho)_\text{HS}\geq0 
\end{align}
for all $\ket{\psi}\in H$ (not necessarily normalized).
These conditions of Hermiticity, unit trace, and positive semi-definiteness, are the probability conditions of reality, normalisation, and non-negativity, placed upon the eigenvalues of $\rho$.
The space is convex in the sense that for any $\rho,\sigma\in D( H)$ and any $a\in[0,1]$, the convex combination $a\rho+(1-a)\sigma$ is also a member of $ D( H)$.
The extreme points of $D( H)$ (that is, densities $\rho$ that may not be formed through a convex combination of distinct densities with $a\in(0,1)$) is the set of pure states $ P( H)$.
These are projections, $\rho^2=\rho$, of rank 1 (as for projections rank is equal to trace), and so satisfy $(\rho,\rho)_\text{HS}=\tr(\rho^2)=\tr(\rho)=1$.
To distinguish pure states from other densities (mixed states) we shall generally express these as outer-products $\ketbra{\psi}{\psi}$, notwithstanding the lack of injectivity in the map $\ket{\psi}\to\ketbra{\psi}{\psi}$.

Adapted to this context, Wigner's theorem states:
\newtheorem*{wigner}{Theorem (Wigner)}
\begin{wigner}
A bijection $W: P( H)\to P( H)$ that satisfies
\begin{align}\label{wigner_condition}
\left(W[\ketbra{\psi}{\psi}],W[\ketbra{\phi}{\phi}]\right)_\text{HS}
=\left(\ketbra{\psi}{\psi},\ketbra{\phi}{\phi}\right)_\text{HS}
\end{align}
for all pure states $\ketbra{\psi}{\psi}$ and $\ketbra{\phi}{\phi}$, is unitary or antiunitary\footnote{Anti-unitary transformations, $\rho\to U\rho^* U^\dagger=U\rho^T U^\dagger$, are related to time reversal, and by extension CPT-symmetry \cite{streaterPCTSpinStatistics1989}.}.
\end{wigner}
Although this is likely the most cited symmetry theorem, the strength of condition \eqref{wigner_condition} is not always acknowledged. 
For instance, we may take $\ketbra{\phi}{\phi}$ to be a pure state $\ketbra{n}{n}$ constructed from an element of an orthonormal Hilbert space basis $\ket{n}$, $n=1,..,N$, so that the RHS of Eq.~\eqref{wigner_condition} becomes the Born probability density $\left(\ketbra{\psi}{\psi},\ketbra{n}{n}\right)_\text{HS}=|\braket{n}{\psi}|^2$.
Condition \eqref{wigner_condition} then entails the conservation of the entire structure of the Born density, not mere overall probability.
In contrast Kadison's theorem (also adapted to this context) states: 
\newtheorem*{kadison}{Theorem (Kadison)}
\begin{kadison}
A bijection from the convex set of density matrices to itself, $K: D( H)\to D( H)$, that conserves the convex structure,
\begin{align}\label{conservation_of_convexity_kadison}
K\left[a\rho + (1-a)\sigma\right]=aK[\rho]+(1-a)K[\sigma]
\end{align}
with $a\in[0,1]$, is unitary or antiunitary.
\end{kadison}
In both Wigner's and Kadison's theorems (as well as the four other symmetry theorems covered in Chp.~5 of Ref.~\cite{landsmanFoundationsQuantumTheory2017}), invertibility is taken to be a condition from which (anti)unitarity is obtained. 
In contrast our theorem (which most closely resembles Kadison's theorem) replaces invertibility with conservation of von Neumann entropy
\begin{align}\label{vNentropy}
S[\rho]=-\tr\rho\log\rho=-\sum_i p_i\log p_i,
\end{align}
where $p_i$ are the eigenvalues of $\rho$. This theorem, the proof of which shall entail the majority of this paper, may be stated in analogy with Kadison's theorem as follows.
\begin{theorem}\label{theorem1}
A map from the convex set of density matrices to itself, $A: D( H)\to D( H)$, that conserves the convex structure,
\begin{align}\label{conservation_of_convexity}
A\left[a\rho + (1-a)\sigma\right]=aA[\rho]+(1-a)A[\sigma],
\end{align}
and that conserves von Neumann entropy $S\left[A[\rho]\right]=S[\rho]$, is unitary or antiunitary.
\end{theorem}


\section{Mathematical preliminaries}
\subsection{Linearity and Sudarshan's A matrices}\label{sec:lin_and_A_matrices}
The requirement to conserve convex combinations, Eq.~\eqref{conservation_of_convexity}, which may be understood as expressing the freedom in how ensembles are defined, would usually constrain map $A$ to be affine.
However, it was noted as far back as 1961 by Sudarshan and collaborators \cite{sudarshanStochasticDynamicsQuantumMechanical1961,jordanDynamicalMappingsDensity1961} that, taken with the additional requirements to conserve probability conditions \eqref{density_matrix_conditions}, $A$ is further constrained to be linear. 
To see this, first note that, using the implied-basis component notation of Ref.~\cite{sudarshanStochasticDynamicsQuantumMechanical1961}, such an affine transformation $\rho\to\rho'$ may be expressed
\begin{align} 
\rho'_{rs}=\sum_{nm}C_{rs;nm}\rho_{nm}+D_{rs}.
\end{align} 
Then, by observing that the trace condition $\sum_{n}\rho_{nn}=1$, means we can write the inhomogeneous term $D_{rs}=\sum_{nm}D_{rs}\delta_{nm}\rho_{nm}$, this can be absorbed into the homogeneous term so that 
\begin{align}
\rho'_{rs}=\sum_{nm}A_{rs;nm}\rho_{nm},
\end{align}
where $A_{rs;nm}=C_{rs;nm}+D_{rs}\delta_{nm}$.
In fact this observation may be made more generally, and applies also in classical contexts. 

In this quantum context, $A_{rs;nm}$ are the elements of \emph{Sudarshan's A matrices} \cite{sudarshanStochasticDynamicsQuantumMechanical1961}.
As we shall be dealing with a number of different bases, we shall instead prefer to suppress indices wherever convenient.
Our notation is related to Sudarshan's component form by
\begin{align}\label{our_A_notation}
\left(A\left[\ketbra{n}{m}\right]\right)_{rs}
=\bra{r}\left(A\left[\ketbra{n}{m}\right]\right)\ket{s}
=A_{rs;nm}.
\end{align}
Sudarshan's A matrices may be thought of as a quantum equivalent to Markov transition matrices, and obey similar conditions to ensure that physical probabilities (density matrix eigenvalues) remain real, normalized, and non-negative. 
Necessary and sufficient conditions for these are respectively 
\begin{align}
A[\ketbra{n}{m}]=A[\ketbra{m}{n}]^\dagger,
\label{q:conservation_of_reality_of_probabilities}\\
\tr\left(A[\ketbra{n}{m}]\right)=\delta_{nm},
\label{q:conservation_of_total_probability}\\
\left(\ketbra{\psi}{\psi},A[\ketbra{\phi}{\phi}]\right)_\text{HS} \geq 0,
\label{q:conservation_of_nonnegative_probabilities}
\end{align}
for all $\ket{\psi},\ket{\phi}\in H$ (not necessarily normalized).
These are proved for instance in Ref.~\cite{jordanDynamicalMappingsDensity1961}, however to help the reader navigate our notation in relation to that of Refs.~\cite{sudarshanStochasticDynamicsQuantumMechanical1961,jordanDynamicalMappingsDensity1961}, we repeat these proofs in Appendix \ref{appendix_nec_suf_for_A}.

We note in passing that, in contrast to the conventional operator-sum representation for quantum operations \cite{nielsenQuantumComputationQuantum2010}, such maps are only positive, not completely positive.
The physical reasonableness of completely positive maps is widely accepted \cite{nielsenQuantumComputationQuantum2010}, though sometimes disputed \cite{pechukasReducedDynamicsNeed1994,shajiWhosAfraidNot2005}.
Our proof does not assume the stronger condition of complete positivity, although this is not by design.
It is simply a result of following, \textit{mutatis mutandis}, the same strategy that was applied to classical contexts \cite{underwoodSignaturesRelicQuantum2019,underwoodPrincipleInformationConservation2020}.


\subsection{von Neumann entropy essentials}
To express von Neumann entropy $S=-\tr\left(\rho\log\rho\right)$ in terms of the eigenvalues $p_i$ of density matrix $\rho$, first recall that as $\rho$ is Hermitian, by the spectral theorem it may be expressed $\rho = X\Lambda X^\dagger$, where $X$ is a unitary matrix with columns equal to normalized eigenvectors of $\rho$, and $\Lambda=\text{diag}(p_1,...,p_N)$. 
For the case that all $p_i$ are positive, the logarithm of $\rho$ is expressible as $\log\rho = X(\log\Lambda)X^\dagger$, where $\log\Lambda=\text{diag}(\log p_1,...,\log p_N)$.
Then, using the cyclic property of the trace, $\tr(ABC)=\tr(CAB)$,
\begin{align}
S&=-\tr\left(\rho\log\rho\right)
=-\tr\left(X\Lambda X^\dagger X(\log\Lambda)X^\dagger\right)\nonumber\\
&=-\tr\left(\Lambda \log\Lambda\right)
=-\sum_i p_i\log p_i.
\end{align}
To see that this is conserved by unitary evolution, $\rho\to \rho'=U\rho U^\dagger$ (i.e., the sufficiency counterpart to our necessity theorem), note that $\rho'=UX\Lambda X^\dagger U^\dagger$, so that 
\begin{align}
S[\rho']&=-\tr\left(UX\Lambda X^\dagger U^\dagger U X(\log\Lambda)X^\dagger U^\dagger\right)\nonumber \\
&=-\tr \left(\Lambda\log\Lambda\right)=S[\rho].
\end{align}

Pure states possess a number of equivalent defining properties \cite{hallQuantumTheoryMathematicians2013,peresQuantumTheoryConcepts2002}; unit purity, $(\rho,\rho)_\text{HS}=\tr(\rho^2)=1$; their status as the extremal points of the convex set of density matrices; idempotency, i.e.,~they are projections. 
One of these is the property of corresponding to zero (minimum) entropy, $S[\rho]=0$.
To see this note that as function $-x\log x$ is strictly positive on the open interval $(0,1)$, is zero for $x=1$, and we follow the standard convention that $0\log 0=0$, justified through the limit $\lim_{x\to 0^+}x\log x=0$, the minimum value entropy \eqref{vNentropy} may take is zero.
This minimum value may only be obtained while retaining normalisation if all eigenvalues $p_i$ vanish except for a single instance of unity. 
Thus all density matrices of zero entropy are unitarily diagonalizable as $\rho=U\Lambda_0U^\dagger$, where $\Lambda_0=\text{diag}(1,0,0...)$.
As $\Lambda_0$ is itself idempotent, $\Lambda_0^2=\Lambda_0$, the idempotency of $\rho$ follows.
So each density matrix of zero entropy is a projection onto an eigenspace corresponding to its unit eigenvalue.
This projection is orthogonal due to the Hermiticity and the eigenspace is one-dimensional due to the unit trace.
Conversely, any one-dimensional orthogonal projection is clearly unitarily similar to $\text{diag}(1,0,0...)$, and thus has zero entropy.

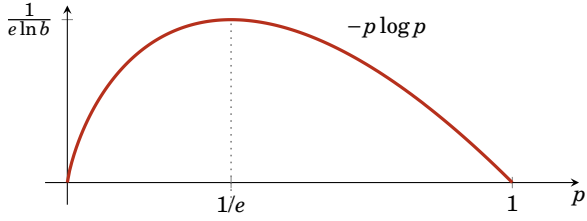
\begin{figure}[!htbp] 
\centering
\begin{tikzpicture}
  \begin{axis}[
    width=246pt,
    height=120pt,
    trim axis left,
    axis lines=middle,
    xlabel={$p$},
    xlabel style={below},
    xmin=-0.05, xmax=1.15,
    ymin=-0.05, ymax=0.4,
    domain=0.00001:0.99999,
    samples=200,
    xtick={0,0.368,1},
    ytick={0,0.368},
    yticklabels={$0$, $\frac{1}{e\ln b}$},
    xticklabels={$0$,$1/e$, $1$},
  ]

    \addplot[BrickRed, line width=1.2pt] {
      -x*ln(x) 
    };

    \node[
      fill=white,
      inner sep=1pt,
      anchor=west
    ] at (axis cs:0.62,0.35) {
      $\displaystyle
      -p\log p$
    };

      \addplot[
        gray,
        dotted,
        line width=0.6pt,
        domain=-0.05:0.368
      ] coordinates {
        (0.368,-0.05)
        (0.368,0.368)
      };
  \end{axis}
\end{tikzpicture}
\caption{Function $-p\log p$ is strictly positive on the open unit interval, $0<p<1$. 
It is zero for $p=1$, and we follow the standard convention that $-0\log0=0$, justified through the limit $\lim_{p\to 0^+}p\log p=0$. 
Our analysis is independent of logarithm base $b$, which in this context merely scales entropy \eqref{vNentropy} by an arbitrary factor.
We leave $b$ unspecified but assume $b>1$ for the purposes of discussion, to retain the conventional interpretations of high and low entropy.  
}\label{fig:xlogx}
\end{figure}

As pure states are uniquely low in entropy, for $A$ to conserve entropy it must map pure states to pure states; for all normalized $\ket{\psi}\in H$ there must exist a corresponding normalized $\ket{\psi'}\in H$ (determined up to a complex phase) such that
\begin{align}\label{q:conservation_of_min_entropy}
\ketbra{\psi'}{\psi'}=A\left[\ketbra{\psi}{\psi}\right].
\end{align}

Maximum entropy, $S=\log N$, corresponds to eigenvalues $p_1=...=p_N=1/N$.
Intuitively this is reflected in the principle of indifference; in the absence of information, no state is special.
To prove $\log N$ is maximal we may apply Jensen's inequality.
This states, for a convex combination ($\sum_i a_i=1$, $a_i\geq0$) of a convex function $f:\mathbb{R}\to\mathbb{R}$, that 
\begin{align}
\sum_i a_if(x_i)\geq f\left(\sum_i a_ix_i\right).
\end{align}
Taking $a_i\to p_i$, $f\to \log(x_i)$ (which is concave, reversing the inequality), and $x_i\to 1/p_i$, $p_i>0$, we find
\begin{align}
S=-\sum_i p_i\log p_i = \sum_i p_i\log\left(\frac{1}{p_i}\right)
\leq \log \left(\sum_i p_i\frac{1}{p_i}\right)
=\log N.
\end{align}
To show that $p_1=...=p_N=1/N$ is the only probability density for which $S=\log N$ is obtained, first note that $S[\rho]$ is strictly concave on the space of eigenvalues (i.e., $\partial^2 S/\partial p_i^2<0$ for all $p_i\in(0,1]$).
Denoting a vector in this eigenvalue space $\mathbf{p}=(p_1,...,p_N)$, this means that on the \textit{open} interval $a\in(0,1)$, a convex combination of any two such vectors $\mathbf{p}^{A}$ and $\mathbf{p}^{B}$, $\mathbf{p}=a\mathbf{p}^{A}+(1-a)\mathbf{p}^{B}$ satisfies $S[\mathbf{p}]>aS[\mathbf{p}^{A}]+(1-a)S[\mathbf{p}^{B}]$ for all $\mathbf{p}^A\neq\mathbf{p}^B$.
Supposing there did exist $\mathbf{p}^{A}\neq\mathbf{p}^{B}$ such that $S[\mathbf{p}^{A}]=S[\mathbf{p}^{B}]=\log N$, then for such a convex combination this would imply $S[a\mathbf{p}^{A} + (1-a)\mathbf{p}^{B}]> \log N$, which is clearly a contradiction.
Thus any density matrix $\rho$ such that $S[\rho]=\log N$ must have eigenvalues $p_1=...=p_N=1/N$.
By the spectral theorem, any density matrix for which this is the case must be unitarily diagonalizable to the maximally mixed state, $\rho_\text{MMS}:=\frac{1}{N}\sum_n\ketbra{n}{n}=I/N$.
Of course, $\rho_{MMS}$ is invariant under unitary transformations, so it is clearly the only density matrix for which $S[\rho]=\log N$. 

As the maximally mixed state is uniquely high in entropy, it must be invariant under any entropy conserving transformation, so that
\begin{align}\label{q:conservation_of_max_entropy}
\sum_n A\left[\ketbra{n}{n}\right]=I.
\end{align}

\subsection{Matrix spaces and useful bases for them}
The space of linear operators $L(H)$ is an $N^2$-dimensional vector space over $\mathbb{C}$.
Given an orthonormal Hilbert space basis $\ket{n}$ for $n=1,...,N$, it is most naturally spanned by bases $\ketbra{n}{m}$ with $n,m=1,...,N$.
These are orthonormal with respect to the Hilbert-Schmidt product,
\begin{align}
\left(\ketbra{n}{m},\ketbra{p}{q}\right)_\text{HS}=\tr(\ket{m}\braket{n}{p}\bra{q})=\delta_{np}\delta_{mq}.
\end{align}
In contrast, the Hermitian matrices $L_\text{sa}(H)\subset L(H)$ are an $N^2$-dimensional vector space over $\mathbb{R}$. 
(A Hermitian matrix multiplied by an imaginary number is no longer Hermitian).
These are most naturally spanned by the basis
\begin{align}
\begin{cases}
\ketbra{n}{n}& n=1,...,N\\
\frac{1}{\sqrt{2}}\left(\ket{m}\bra{n}+\ketbra{n}{m}\right)& m> n=1,...,N\\
\frac{i}{\sqrt{2}}\left(\ketbra{m}{n}-\ketbra{n}{m}\right)& m>n=1,...,N
\end{cases},
\end{align}
which is also orthonormal with respect to the Hilbert-Schmidt product.
Density matrices $D(H)\subset L_\text{sa}(H)$ lose one dimension due to the trace condition \eqref{density_matrix_conditions}, and are no longer a true vector space as the trace condition means the space is no longer closed under scalar multiplication or vector addition. 
Pure states $P(H)\subset D(H)$ occupy a much smaller space, of dimension $2(N-1)$ \cite{grabowskiGeometryQuantumSystems2005}.
Nevertheless, the normalized but non-orthogonal set of pure states,
\begin{align}\label{pure_state_basis}
\begin{cases}
\ketbra{n}{n}& n=1,...,N\\
\frac{1}{\sqrt{2}}\left(\ket{n}+\ket{m}\right)\frac{1}{\sqrt{2}}\left(\bra{n}+\bra{m}\right)& m> n=1,...,N\\
\frac{1}{\sqrt{2}}\left(\ket{n}+i\ket{m}\right)\frac{1}{\sqrt{2}}\left(\bra{n}-i\bra{m}\right)& m>n=1,...,N
\end{cases},
\end{align}
may through real combinations, serve as a basis for $L_\text{sa}(H)$, or through complex combinations, serve as a basis for $L(H)$.
Much of the below proof involves deducing properties of the transformation of individual basis elements for $L(H)$, $\ketbra{m}{n}$, from those of the (zero entropy) pure states \eqref{pure_state_basis}.


\section{Proof of theorem 1}
Although we do not claim to approach the structural rigor that might be demanded in some mathematical circles, our proof (is not elegant, and so) benefits from breaking up into parts. 
We present these as seven separate propositions, with associated individual proofs.
To help indicate our strategy to the reader, the final piece of the puzzle is stated upfront as Proposition \hyperlink{thm0}{0} (in a nod to its thermodynamical motivation).
This states that, given some orthonormal $H$ basis $\ket{n}$, if there is a corresponding primed orthonormal basis $\ket{n''}$ such that linear\footnote{Linearity is established as a consequence of the conservation of convexity \eqref{conservation_of_convexity} and probability conditions \eqref{density_matrix_conditions} in Sec.~\ref{sec:lin_and_A_matrices}.} transformation $A$ maps $L(H)$ basis elements as $\ketbra{n}{m}\to\ketbra{n''}{m''}$, then $A$ is unitary.
(The double-primes are used for later convenience.)
The purpose of Propositions \hyperlink{thm1}{1}-\hyperlink{thm7}{7} is to establish that this is indeed the case for an $A$ that conserves von Neumann entropy.
The possibility of an antiunitary $A$ arises as a second solution, corresponding to $\ketbra{n}{m}\to\ketbra{m''}{n''}$.
The key insight that enables the rest of the proof to follow is contained in Proposition \hyperlink{thm1}{1}.
This shows that this is true for ``diagonal elements'', $\ketbra{n}{n}\to\ketbra{n''}{n''}$, while Propositions \hyperlink{thm2}{2}-\hyperlink{thm7}{7} extend this to the off-diagonals. 


\newtheorem*{thm0}{Proposition 0}
\begin{thm0}\hypertarget{thm0}{}
For $n,m=1,...,N$, let $\ketbra{n}{m}$ and $\ketbra{n''}{m''}$ be the basis vectors of two orthonormal bases of $ L( H)$.
A linear transformation on the space of density matrices, $A:D(H)\to D(H)$, that maps basis vectors as $\ketbra{n}{m}\to\ketbra{n''}{m''}$ is unitary, $A[\rho]=U\rho U^\dagger$ (that is, where $U$ is a unitary matrix).
Similarly, a linear transformation that maps basis vectors as $\ketbra{n}{m}\to\ketbra{m''}{n''}$ is antiunitary, $A[\rho]=U\rho^* U^\dagger=U\rho^T U^\dagger$.
\end{thm0}


\theoremstyle{definition}
\newtheorem*{prf0}{Proof of Proposition \hyperlink{thm0}{0}}
\begin{prf0}
First consider that 
\begin{align}
\ketbra{n''}{m''}
&=\left(\sum_r\ketbra{r''}{r}\right)\ketbra{n}{m}\left(\sum_s\ketbra{s}{s''}\right)
=U\ketbra{n}{m}U^\dagger,
\end{align}
where sums over the (double-)primed vectors are implied and we have defined $U:=\sum_r\ketbra{r''}{r}$. The unitarity of this matrix is clear,
\begin{align}
UU^\dagger=\sum_{rs}\ket{r''}\braket{r}{s}\bra{s''}
&=\sum_r\ketbra{r''}{r''}=I,\\
U^\dagger U=\sum_{rs}\ket{s}\braket{s''}{r''}\bra{r}
&=\sum_s\ketbra{s}{s}=I.
\end{align}
As any matrix is expandable as $\rho=\sum_{nm}\rho_{nm}\ketbra{n}{m}$, if $A[\ketbra{n}{m}]=\ketbra{n''}{m''}$, linearity ensures that $A[\rho]=U\rho U^\dagger$.

For the case that $A[\ketbra{n}{m}]=\ketbra{m''}{n''}$, an arbitrary matrix transforms as
\begin{align}
A\left[\rho\right]=\sum_{nm}\rho_{nm}\ketbra{m''}{n''}
=\sum_{nm}\rho_{mn}\ketbra{n''}{m''}
=U\rho^T U^\dagger,
\end{align}
which is equal to $U\rho^* U^\dagger$, due to the Hermiticity of $\rho$.\qed
\end{prf0}

We now impose conservation of entropy and attempt to arrive at the input of Proposition \hyperlink{thm0}{0}.


\newtheorem*{thm1}{Proposition 1}
\begin{thm1}\hypertarget{thm1}{}
Let $\ket{n}$, $n=1,...,N$ be an orthonormal $ H$ basis, and let $A$ be a linear transformation on density matrices $A: D(H)\to D(H)$ that conserves entropy, $S\left[A[\rho]\right]=S[\rho]$. 
Then there exists a corresponding ``primed'' orthonormal $H$ basis $\ket{n'}$, $n=1,...,N$ such that diagonal $ L( H)$ basis elements are mapped as $A\left[\ketbra{n}{n}\right]=\ketbra{n'}{n'}$.
\end{thm1}
\theoremstyle{definition}
\newtheorem*{prf1}{Proof of Proposition \hyperlink{thm1}{1}}
\begin{prf1}
From the conservation of minimum entropy condition Eq.~\eqref{q:conservation_of_min_entropy}, $A$ must map pure states to pure states so that for all $n$ there is a pure state $\ketbra{n'}{n'}$ such that $\ketbra{n'}{n'}=A\left[\ketbra{n}{n}\right]$.
As these pure states $\ketbra{n'}{n'}$ are invariant under $\ket{n'}\to e^{i\theta}\ket{n'}$, this does not uniquely specify a corresponding set $\ket{n'}$, $n=1,...,N$.
Our task is to show that any such corresponding set of $\ket{n'}$ used to construct $\ketbra{n'}{n'}$, are orthonormal $\braket{n'}{m'}=\delta_{nm}$.
If this is so, they will naturally span $H$.

To demonstrate orthonormality, first note that applied to density matrices, the Hilbert-Schmidt norm corresponds to the standard test of purity $\tr(\rho^2)$, so that  
\begin{align}
\left(\ketbra{n'}{n'},\ketbra{n'}{n'}\right)_\text{HS}=1.
\end{align}
Consider also the sum of products
\begin{align}
&\sum_m\left(\ketbra{n'}{n'},\ketbra{m'}{m'}\right)_\text{HS}\nonumber\\
=&\left(A\left[\ketbra{n}{n}\right],\sum_mA\left[\ketbra{m}{m}\right]\right)_\text{HS}\\
=&\left(A\left[\ketbra{n}{n}\right],I\right)_\text{HS}\label{q:prop1_prf1}\\
=&\tr\left(A\left[\ketbra{n}{n}\right]^\dagger\right)=\delta_{nn}=1,\label{q:prop1_prf2}
\end{align}
where sums over primed vectors are implied, to reach Eq.~\eqref{q:prop1_prf1} we have used the conservation of max entropy property \eqref{q:conservation_of_max_entropy}, and to reach Eq.~\eqref{q:prop1_prf2} we have used conservation of both the reality of probabilities \eqref{q:conservation_of_reality_of_probabilities} and total probability \eqref{q:conservation_of_total_probability}.

These Hilbert-Schmidt products on $L(H)$ vectors may be recast in terms of inner products on $H$ vectors, so that both 
\begin{align}
\sum_m\left(\ketbra{n'}{n'},\ketbra{m'}{m'}\right)_\text{HS}
&=\sum_m\left|\braket{n'}{m'}\right|^2=1,
\end{align}
and
\begin{align}
\left(\ketbra{n'}{n'},\ketbra{n'}{n'}\right)_\text{HS}
&=\left|\braket{n'}{n'}\right|^2=1.
\end{align}
Hence $\left|\braket{n'}{m'}\right|^2=\delta_{nm}$ for all $n,m$, and as
$\braket{n'}{n'}^*=\braket{n'}{n'}$, and $\left|\braket{n'}{m'}\right|^2=0$ if and only if $\braket{n'}{m'}=0$, we conclude $\braket{n'}{m'}=\delta_{nm}$.
\qed
\end{prf1}


A straightforward corollary to this is that, given such a basis $\ket{n'}$, the outer products $\ketbra{n'}{m'}$ ($m,n=1,...,N$) satisfy 
$\left(\ketbra{n'}{m'},\ketbra{p'}{q'}\right)_\text{HS}=\delta_{np}\delta_{mq}$ and so form an orthonormal basis for $ L( H)$.

To establish unitarity, Proposition \hyperlink{thm0}{0} requires $\ketbra{n}{m}\to\ketbra{n'}{m'}$ for all $n,m$, not merely the diagonals $n=m$ established in Proposition \hyperlink{thm1}{1}.
Propositions \hyperlink{thm2}{2}-\hyperlink{thm7}{7} isolate how the off-diagonals transform into the primed $ L( H)$ basis.


\newtheorem*{thm2}{Proposition 2}
\begin{thm2}\hypertarget{thm2}{}
Given a primed Hilbert space basis $\ket{r'}$, defined such that $\ketbra{r'}{r'}=A\left[\ketbra{r}{r}\right]$ as described above, then for any pair of distinct basis vectors $\ket{n}$ and $\ket{m}$, 
\begin{align}
A\left[\ketbra{n}{m}+\ketbra{m}{n}\right]
=e^{i(\alpha-\beta)}\ketbra{n'}{m'} + e^{-i(\alpha-\beta)}\ketbra{m'}{n'}\label{alphabeta}
\end{align}
for some angles $\alpha$ and $\beta$.
\end{thm2}
We split this into two parts as follows.
\newtheorem*{thm2a}{Proposition 2a}
\begin{thm2a}\hypertarget{thm2a}{}
For some $-1\leq a \leq 1$ and some angles $\alpha$, $\beta$,
\begin{align}
&A\left[\ketbra{n}{m}+\ketbra{m}{n}\right]\nonumber\\
=&
a\ketbra{n'}{n'}-a\ketbra{m'}{m'}\nonumber\\
&+\sqrt{(1+a)(1-a)}e^{i(\alpha-\beta)}\ketbra{n'}{m'}\nonumber\\
&+\sqrt{(1+a)(1-a)}e^{-i(\alpha-\beta)}\ketbra{m'}{n'}.\label{prp2a}
\end{align}
\end{thm2a}


\theoremstyle{definition}
\newtheorem*{prf2a}{Proof of Proposition \hyperlink{thm2a}{2a}}
\begin{prf2a}
We'll consider how the sum of off-diagonals $\ketbra{n}{m}+\ketbra{m}{n}$ is transformed as part of the pure states corresponding to quantum states $\ket{\psi_+}=\frac{1}{\sqrt{2}}\left(\ket{n}+\ket{m}\right)$ and $\ket{\psi_-}=\frac{1}{\sqrt{2}}\left(\ket{n}-\ket{m}\right)$.
From Proposition \hyperlink{thm1}{1},
\begin{align}
&A\left[\ketbra{\psi_\pm}{\psi_\pm}\right]\\
=&A\left[\frac{1}{\sqrt{2}}\left(\ket{n}\pm\ket{m}\right)\frac{1}{\sqrt{2}}\left(\bra{n}\pm\bra{m}\right)\right]\\
=&A\left[\frac{1}{2}\left(\ketbra{n}{n}+\ketbra{m}{m}\pm\ketbra{n}{m}\pm\ketbra{m}{n}\right)\right]\\
=&\frac{1}{2}\left(\ketbra{n'}{n'}+\ketbra{m'}{m'}\pm A\left[\ketbra{n}{m}+\ketbra{m}{n}\right]\right).\label{prp2a1}
\end{align}
Denote the diagonal components of $A\left[\ketbra{n}{m}+\ketbra{m}{n}\right]$ in the primed $ L( H)$ basis
$a_r:=\bra{r'}A\left[\ketbra{n}{m}+\ketbra{m}{n}\right]\ket{r'}$.
From Hermiticity of matrix \eqref{prp2a1}, it follows that $a_1,...,a_N\in\mathbb{R}$.
From positive semi-definiteness with respect to the primed basis, $\bra{r'}A\left[\ketbra{\psi_\pm}{\psi_\pm}\right]\ket{r'}\geq0$, it follows that $\pm a_r\geq0$ for all $r\neq n,m$, and $\frac{1}{2}(1\pm a_r)\geq0$ for $r=n,m$.
From the unit trace property, $1+\frac{1}{2}\sum_r a_r=1$.
Together, these properties mean that $a_r=0$ for all $r\neq n,m$, and there is some $a\in[-1,1]$ such that $a_n=-a_m=a$.

Now, as \eqref{prp2a1} is a pure state, it may be expanded 
$A\left[\ketbra{\psi_+}{\psi_+}\right]
=\sum_{rs}(\psi_+')_r\overline{(\psi_+')}_s\ketbra{r'}{s'}$.
We may identify its diagonal elements as 
$\left|(\psi_+')_n\right|^2=\frac{1}{2}(1+a)$,  
$\left|(\psi_+')_m\right|^2=\frac{1}{2}(1-a)$, and
$\left|(\psi_+')_r\right|^2=0$ for all $r\neq n,m$.
Hence there are two angles $\alpha,\beta$ such that 
$(\psi_+')_n=\sqrt{(1+a)/2}e^{i\alpha}$ and 
$(\psi_+')_m=\sqrt{(1-a)/2}e^{i\beta}$,
meaning we may reexpress the state
\begin{align}
A\left[\ketbra{\psi_+}{\psi_+}\right]
=&\frac{1}{2}\large[
(1+a)\ketbra{n'}{n'}
+(1-a)\ketbra{m'}{m'}\nonumber\\
&+\sqrt{(1+a)(1-a)}e^{i(\alpha-\beta)}\ketbra{n'}{m'}\nonumber\\
&+\sqrt{(1+a)(1-a)}e^{-i(\alpha-\beta)}\ketbra{m'}{n'}
\large].\label{prp2a2}
\end{align}
Eq.~\eqref{prp2a} is found by isolating $A\left[\ketbra{n}{m}+\ketbra{m}{n}\right]$ from the equality of Eqs.~\eqref{prp2a1} and \eqref{prp2a2}.
\qed
\end{prf2a}


\newtheorem*{thm2b}{Proposition 2b}
\begin{thm2b}\hypertarget{thm2b}{}
In Eq.~\eqref{prp2a}, $a=0$, so that it reduces to the result of Proposition \hyperlink{thm2}{2}, Eq.~\eqref{alphabeta}.
\end{thm2b}


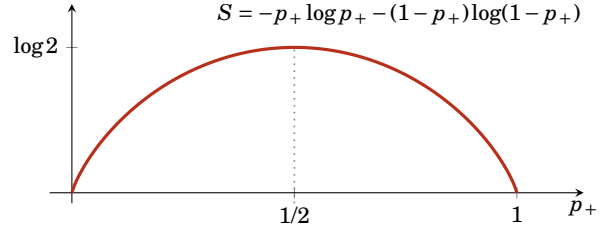
\begin{figure}[!htbp] 
\centering
\begin{tikzpicture}
  \begin{axis}[
    width=246pt,
    height=120pt,
    trim axis left,
    axis lines=middle,
    xlabel={$p_{+}$},
    xlabel style={below},
    xmin=-0.05, xmax=1.15,
    ymin=-0.05, ymax=0.90,
    domain=0.00001:0.99999,
    samples=200,
    xtick={0,0.5,1},
    ytick={0,0.693},
    yticklabels={$0$, $\log 2$},
    xticklabels={$0$,$1/2$, $1$},
  ]

    \addplot[BrickRed, line width=1.2pt] {
      -x*ln(x) - (1-x)*ln(1-x)
    };

    \node[
      fill=white,
      inner sep=1pt,
      anchor=west
    ] at (axis cs:0.32,0.85) {
      $\displaystyle
      S=-p_{+}\log p_{+}
      -(1-p_{+})\log(1-p_{+})$
    };

      \addplot[
        gray,
        dotted,
        line width=0.6pt,
        domain=-0.05:0.693
      ] coordinates {
        (0.5,-0.05)
        (0.5,0.693)
      };
  \end{axis}
\end{tikzpicture}
\caption{The proofs of Propositions 2b-4 consider density matrices for which all eigenvalues/probabilities $p_i$ vanish except for two that we refer to as $p_+$ and $p_-=1-p_+$.
As we retain the standard $0\log0=0$ convention, the von Neumann entropy in this case is the single variable function $S=-p_+\log p_+-(1-p_+)\log(1-p_+)$, which is symmetric about a maximum of $\log2$ at $p_+=p_-=\frac{1}{2}$, and minima of zero at $p_+=0$ or 1.}\label{fig:two_component_entropy}
\end{figure}

\theoremstyle{definition}
\newtheorem*{prf2b}{Proof of Proposition \hyperlink{thm2b}{2b}}
\begin{prf2b}
Consider now the mixed state
\begin{align}
\rho=&\frac{1}{2}\ketbra{n}{n}+\frac{1}{2}\ketbra{\psi_+}{\psi_+}\nonumber\\
=&\frac{3}{4}\ketbra{n}{n}+\frac{1}{4}\left(\ketbra{m}{m}+\ketbra{n}{m}+\ketbra{m}{n}\right).
\end{align}
To calculate the entropy of this state, we first look to find its eigenvalues $p_i$, and so solve the characteristic equation, $0=\det(\rho-p_i I)$.
Matrix $\rho-p_i I$ may be put into block diagonal form by switching row 1 with row $n$, column 1 with column $n$, row 2 with row $m$, and column 2 with column $m$.
The resulting characteristic equation is
\begin{align}
0=\det
\begin{pmatrix}
A&B\\C&D
\end{pmatrix}
\text{, where }\ \ 
A=
\begin{pmatrix}
\frac{3}{4}-p_i & \frac{1}{4}\\
\frac{1}{4} & \frac{1}{4}-p_i
\end{pmatrix},
\end{align}
$B$ and $C$ are null, and $D=-p_i I_{N-2}$ where $I_{N-2}$ is the $(N-2)$-dimensional identity.
The determinant of such block diagonal matrices is equal to $\det(A-BD^{-1}C)\det{D}$ provided $D$ is invertible.
(In this case $D^{-1}=-I_{N-2}/p_i$, which exists if $p_i\neq0$).
The resulting characteristic equation is $0=(-p_i)^{N-2}\left(p_i^2-p_i+\frac{1}{8}\right)$, which has the nontrivial ($p_i\neq0$) solutions
$p_\pm=\frac{1}{2}\pm\frac{1}{\sqrt{8}}$.
As these sum to unity, we may conclude that the remaining eigenvalues all vanish.

The same process may be followed for the transformed matrix
\begin{align}
A[\rho]
=&\frac{3}{4}\ketbra{n'}{n'}+\frac{1}{4}\left(\ketbra{m'}{m'}+A\left[\ketbra{n}{m}+\ketbra{m}{n}\right]\right).
\end{align}
Using expression \eqref{prp2a} results in the characteristic equation $0=(-p_i)^{N-2}\left(p_i^2-p_i+\frac{1}{8}(1-a)\right)$, which has the nontrivial ($p_i\neq0$) solutions $p_\pm=\frac{1}{2}\pm\sqrt{(1+a)/8}$.
Previously, Proposition \hyperlink{thm2a}{2a} established $a\in[-1,1]$. 
Now we see that $-1\leq a < 0$ would bring $p_\pm$ closer to $\frac{1}{2}$, increasing entropy (see Fig.~\ref{fig:two_component_entropy}), 
and $0 < a \leq 1$ would take eigenvalues further from $\frac{1}{2}$ causing a decrease in entropy.
Thus, for $S\left[A[\rho]\right]$ to equal $S\left[\rho\right]$, $a$ must vanish.
\qed
\end{prf2b}


Proposition \hyperlink{thm3}{3} is stated without proof, as this follows in close analogy with that of Propositions \hyperlink{thm2a}{2a} and \hyperlink{thm2b}{2b} with the simple replacement,
\begin{align}
\ket{\psi_\pm}\longrightarrow \frac{1}{\sqrt{2}}\left(\ket{n} \pm i \ket{m}\right).
\end{align}


\newtheorem*{thm3}{Proposition 3}
\begin{thm3}\hypertarget{thm3}{}
Given a Hilbert space basis $\ket{r'}$ such that $\ketbra{r'}{r'}=A\left[\ketbra{r}{r}\right]$ as described above, then for any pair of distinct basis vectors $\ket{n}$ and $\ket{m}$, 
\begin{align}
A\left[-i\ketbra{n}{m}+i\ketbra{m}{n}\right]
=e^{i(\gamma-\delta)}\ketbra{n'}{m'} + e^{-i(\gamma-\delta)}\ketbra{m'}{n'}\label{gammadelta}
\end{align}
for some $\gamma$ and $\delta$.
\end{thm3}


So far, for any particular choice of $n\neq m$, we find Eqs.~\eqref{alphabeta} and \eqref{gammadelta}, relating the pair of off-diagonals $\ket{n}\bra{m}$ and $\ket{m}\bra{n}$ to a corresponding pair of transformed off-diagonals in the primed basis, $\ket{n'}\bra{m'}$ and $\ket{m'}\bra{n'}$.
Our next step is to find a relationship between angles $\alpha$, $\beta$, $\gamma$, and $\delta$, which allows us to isolate a single member of each pair (i.e., telling us how $\ket{n}\bra{m}$ transforms without reference to $\ket{m}\bra{n}$).
In doing so we find a dual solution, which ends up corresponding to the dual conclusions of unitarity and antiunitarity.


\newtheorem*{thm4}{Proposition 4}
\begin{thm4}\hypertarget{thm4}{}
With the above considerations, for any distinct pair $n$ and $m$, there is a corresponding pair of angles $\alpha$ and $\beta$ such that 
\begin{align}
\text{Possibility A:}\quad
\begin{cases}
A\left[\ketbra{n}{m}\right]=e^{i(\alpha-\beta)}\ketbra{n'}{m'}\\
A\left[\ketbra{m}{n}\right]=e^{-i(\alpha-\beta)}\ketbra{m'}{n'}
\end{cases},\label{possA_w_angles}
\end{align}
or
\begin{align}
\text{Possibility B:}\quad
\begin{cases}
A\left[\ketbra{n}{m}\right]=e^{-i(\alpha-\beta)}\ketbra{m'}{n'}\\
A\left[\ketbra{m}{n}\right]=e^{i(\alpha-\beta)}\ketbra{n'}{m'}
\end{cases}\label{possB_w_angles}.
\end{align}
\end{thm4}


\theoremstyle{definition}
\newtheorem*{prf4}{Proof of Proposition \hyperlink{thm4}{4}}
\begin{prf4}
Consider the mixed state $\frac{1}{2}\left(\rho_A + \rho_B\right)$, where 
\begin{align}
\rho_A:=\frac{1}{2}\left(\ketbra{n}{n}+\ketbra{m}{m}+\ketbra{n}{m}+\ketbra{m}{n}\right),\\
\rho_B:=\frac{1}{2}\left(\ketbra{n}{n}+\ketbra{m}{m}-i\ketbra{n}{m}+i\ketbra{m}{n}\right).
\end{align}
Using the same method as previously, this may be shown to possess the eigenvalues $p_\pm=\frac{1}{2}\pm\frac{1}{\sqrt{8}}$, with all others vanishing.
The transformed density matrix $A\left[\frac{1}{2}\left(\rho_A + \rho_B\right)\right]$ on the other hand possesses only the non-zero eigenvalues $p_\pm=\frac{1}{2}\pm\frac{1}{4}|z|$, where $z=e^{i(\alpha-\beta)}+e^{i(\gamma-\delta)}$.
Just as previously, $S$ is of the two-component form shown in Fig.~\ref{fig:two_component_entropy}.
If $\frac{1}{4}|z| < \frac{1}{\sqrt{8}}$, then the transformed values of $p_\pm$ would be closer to one half than originally, causing an increase in entropy. 
If $\frac{1}{4}|z| > \frac{1}{\sqrt{8}}$, a reduction in entropy would result. 
For entropy to be conserved, it must be the case that $\left|z\right|=\sqrt{2}$, so that $(\alpha-\beta)-(\gamma-\delta)= \frac{\pi}{2}+k\pi$ for some $k\in\mathbb{Z}$.
This allows us to relate the exponential factors in Eq.~\eqref{gammadelta} to those in Eq.~\eqref{alphabeta}, 
\begin{align}
e^{i(\gamma-\delta)}
&=e^{i((\alpha-\beta)-(\frac{\pi}{2}+k\pi))}\nonumber
=-ie^{-ik\pi}e^{i(\alpha-\beta)}\\
&=\begin{cases}
-ie^{i(\alpha-\beta)}&(\text{even } k)\\
+ie^{i(\alpha-\beta)}&(\text{odd } k)
\end{cases},\\
e^{-i(\gamma-\delta)}
&=\begin{cases}
+ie^{-i(\alpha-\beta)}&(\text{even } k)\\
-ie^{-i(\alpha-\beta)}&(\text{odd } k)
\end{cases}.
\end{align}
Substitution of each case into \eqref{gammadelta}, gives 
\begin{align}
&A\left[\ketbra{n}{m}-\ketbra{m}{n}\right]\nonumber\\
& = 
\begin{cases}
e^{i(\alpha-\beta)}\ketbra{n'}{m'} - e^{-i(\alpha-\beta)}\ketbra{m'}{n'}&(\text{even }k)\\
-e^{i(\alpha-\beta)}\ketbra{n'}{m'} + e^{-i(\alpha-\beta)}\ketbra{m'}{n'}&(\text{odd }k)
\end{cases}.\label{gammadeltaalphabeta}
\end{align}
Finally, taking half the sum of Eq.~\eqref{gammadeltaalphabeta} with Eq.~\eqref{alphabeta} gives
\begin{align}
A\left[\ketbra{n}{m}\right]=
\begin{cases}
e^{i(\alpha-\beta)}\ket{n'}\bra{m'}&(\text{even }k)\\
e^{-i(\alpha-\beta)}\ket{m'}\bra{n'}&(\text{odd }k)
\end{cases},
\end{align}
and half the difference gives
\begin{align}
A\left[\ketbra{m}{n}\right]=
\begin{cases}
e^{-i(\alpha-\beta)}\ket{m'}\bra{n'}&(\text{even }k)\\
e^{i(\alpha-\beta)}\ket{n'}\bra{m'}&(\text{odd }k)
\end{cases}.
\end{align}
In the statement of Proposition \hyperlink{thm4}{4}, possibility A corresponds to even $k$, and possibility B corresponds to odd $k$.
\qed
\end{prf4}


So far, we have considered only pairs of off-diagonals in isolation from other off-diagonal pairs. 
Thus, for each distinct pair $\ketbra{n}{m}$ and $\ketbra{m}{n}$, there is a potentially distinct pair of angles $\alpha$ and $\beta$ such that Proposition \hyperlink{thm4}{4} holds. 
To distinguish these, for all $m>n$ define angles $\theta_{nm}$ equal to the difference $(\alpha-\beta)$ corresponding to the pair $n,m$.
Then, for all $m>n$, Eq.~\eqref{possA_w_angles} becomes
\begin{align}
\text{Possibility A:}\quad
\begin{cases}
A\left[\ketbra{n}{m}\right]=e^{i\theta_{nm}}\ketbra{n'}{m'}\\
A\left[\ketbra{m}{n}\right]=e^{-i\theta_{nm}}\ketbra{m'}{n'}
\end{cases},\label{notation_possA_w_angles}
\end{align}
and Eq.~\eqref{possB_w_angles} becomes
\begin{align}
\text{Possibility B:}\quad
\begin{cases}
A\left[\ketbra{n}{m}\right]=e^{-i\theta_{nm}}\ketbra{m'}{n'}\\
A\left[\ketbra{m}{n}\right]=e^{i\theta_{nm}}\ketbra{n'}{m'}
\end{cases}\label{notation_possB_w_angles}.
\end{align}
The purpose of Proposition \hyperlink{thm5}{5} is to demonstrate a relationship between the $\frac{1}{2}(N^2-N)$ angles $\theta_{nm}$, and in so doing replace them with the angles $\theta_1,...,\theta_N$.

\newtheorem*{thm5}{Proposition 5}
\begin{thm5}\hypertarget{thm5}{}
There is a set of angles $\theta_1,...,\theta_N$, such that for each pair $n\neq m$,
\begin{align}
&\text{Possibility A:}\quad A\left[\ketbra{n}{m}\right]=e^{i(\theta_n-\theta_m)}\ketbra{n'}{m'},\label{prop5A}\\
&\text{Possibility B:}\quad A\left[\ketbra{n}{m}\right]=e^{-i(\theta_n-\theta_m)}\ketbra{m'}{n'}\label{prop5B}.
\end{align}
\end{thm5}

\theoremstyle{definition}
\newtheorem*{prf5}{Proof of Proposition \hyperlink{thm5}{5}}
\begin{prf5}
Consider the quantum state $\ket{\psi}=\frac{1}{\sqrt{N}}(\ket{1}+...+\ket{N})$.
The corresponding (pure state) density matrix may be expanded as
\begin{align}
\ketbra{\psi}{\psi}
&=\frac{1}{N}\sum_n\ketbra{n}{n}
+\frac{1}{N}\sum_{m>n}\left(\ketbra{n}{m}+\ketbra{m}{n}\right).
\end{align}
By Proposition \hyperlink{thm1}{1}, and Eqs.~\eqref{notation_possA_w_angles} and \eqref{notation_possB_w_angles}, for both Possibility A and B this transforms to
\begin{align}
A\left[\ketbra{\psi}{\psi}\right]
=&\frac{1}{N}\sum_n\ketbra{n'}{n'}\nonumber\\
&+\frac{1}{N}\sum_{m>n}\left(e^{i\theta_{nm}}\ketbra{n'}{m'}+e^{-i\theta_{nm}}\ketbra{m'}{n'}\right),\label{prop5_expansion1}
\end{align}
where the sum over the primed basis vectors is implied.
As this is pure (to conserve minimum entropy) it also corresponds to some quantum state $\ket{\psi'}=\sum_n\psi'_n\ket{n'}$, and so may be expanded 
\begin{align}
&A\left[\ketbra{\psi}{\psi}\right]
=\sum_{nm}\psi'_n\bar{\psi}'_m\ketbra{n'}{m'}\nonumber\\
&=\sum_n|\psi'_n|^2\ketbra{n'}{n'}
+\sum_{m>n}\left(\psi'_n\bar{\psi}'_m\ketbra{n'}{m'}+\psi'_m\bar{\psi}'_n\ketbra{m'}{n'}\right),\label{prop5_expansion2}
\end{align}
(Distinguishing between Possibility A and B here leads to the same conclusion.)
A comparison of the diagonal elements of expressions \eqref{prop5_expansion1} and \eqref{prop5_expansion2} gives $\left|\psi'_n\right|^2=1/N$ for all $n$.
Then by writing $\psi'_n=\frac{1}{\sqrt{N}}e^{i\theta_{n}}$, a comparison of the off-diagonals in the upper triangle ($m>n$) results in $e^{i\theta_{nm}}=e^{i(\theta_n-\theta_m)}$, and in the lower triangle $e^{-i\theta_{nm}}=e^{-i(\theta_n-\theta_m)}$. 
Thus for all basis elements we may make the association $\theta_{nm}=\theta_n-\theta_m$, which when substituted into Eqs.~\eqref{notation_possA_w_angles} and \eqref{notation_possB_w_angles}, gives Eqs.~\eqref{prop5A} and \eqref{prop5B} respectively. 
\qed
\end{prf5}


\newtheorem*{thm6}{Proposition 6}
\begin{thm6}\hypertarget{thm6}{}
There is an $H$ basis, $\ket{1''},...,\ket{N''}$, for which the pairwise options described in Eqs.~\eqref{prop5A} and \eqref{prop5B} simplify to 
\begin{align}
\text{Possibility A:}\quad
&A\left[\ketbra{n}{m}\right]=\ketbra{n''}{m''},\label{prop6A}\\
\text{Possibility B:}\quad
&A\left[\ketbra{n}{m}\right]=\ketbra{m''}{n''}.\label{prop6B}
\end{align}
\end{thm6}


\theoremstyle{definition}
\newtheorem*{prf6}{Proof of Proposition \hyperlink{thm6}{6}}
\begin{prf6}
As noted briefly in the proof of Proposition \hyperlink{thm1}{1}, the single-primed $ H$ basis $\ket{n'}$, defined such that $\ketbra{n'}{n'}=A\left[\ketbra{n}{n}\right]$, is not unique.
Given such a $\ket{n'}$ basis and a set of $N$ angles $\varphi_n$, a new basis $\ket{n'}\to e^{i\varphi_n}\ket{n'}$ also satisfies this property.
To this end, Eqs.~(\ref{prop6A},\ref{prop6B}) correspond to Eqs.~(\ref{prop5A},\ref{prop5B}) expressed in terms of a double-primed basis defined as $\ket{n''}:=e^{i\theta_n}\ket{n'}$. 
\qed
\end{prf6}


At present, Possibilities A and B appear to be pairwise alternatives for each pair $n\neq m$.
Proposition \hyperlink{thm6}{6} establishes that for $N\geq 3$ (where there are more than a single pair of off-diagonals), this is in fact a single global distinction; all pairs $m\neq n$ must ``make the same choice'' of either Possibility A or B.

 
\newtheorem*{thm7}{Proposition 7}
\begin{thm7}\hypertarget{thm7}{}
The distinction between Possibility A and Possibility B is global, and not merely pairwise, so that either all basis elements $\ketbra{n}{m}$, $n,m=1,...,N$, transform as in Eq.~\eqref{prop6A} or all transform as in Eq.~\eqref{prop6B}.
\end{thm7}


\theoremstyle{definition}
\newtheorem*{prf7}{Proof of Proposition \hyperlink{thm7}{7}}
\begin{prf7}
WLOG, let $N=3$ and consider the state vector $\ket{\psi}=\left(\ket{1}+i\ket{2}+\ket{3}\right)/\sqrt{3}$, for which the components of the desnity matrix in the unprimed basis are
\begin{align}
\frac{1}{3}
\begin{pmatrix}
1 & -i & 1 \\
i & 1 & i \\
1 & -i & 1 
\end{pmatrix}.
\end{align}
By Eqs.~(\ref{prop6A},\ref{prop6B}), the components of the transformed density matrix $A\left[\ketbra{\psi}{\psi}\right]$ in the double-primed basis $\ket{n''}$, $n=1,2,3$ are 
\begin{align}
\frac{1}{3}
\begin{pmatrix}
1 & -i\epsilon_{12} & 1 \\
i\epsilon_{12} & 1 & i\epsilon_{23} \\
1 & -i\epsilon_{23} & 1 
\end{pmatrix},
\end{align}
where $\epsilon_{12}=1$ for Possibility A and $\epsilon_{12}=-1$ for Possibility B, and similarly for $\epsilon_{23}$.
As $A\left[\ketbra{\psi}{\psi}\right]$ is pure, it may be expanded $A\left[\ketbra{\psi}{\psi}\right]=\sum_{n,m}\psi''_n\bar{\psi}''_m\ketbra{n''}{m''}$.
From the diagonals, $|\psi''_1|^2=|\psi''_2|^2=|\psi''_3|^2=1/3$.
Writing $\psi''_n=e^{i\phi_n}/\sqrt{3}$, we see from the off-diagonals that
\begin{align}
&(1,2):\qquad e^{i\left(\phi_1-\phi_2\right)} = -i\epsilon_{12},\label{prop7A}\\
&(1,3):\qquad e^{i\left(\phi_1-\phi_3\right)} = 1,\label{prop7B}\\
&(2,3):\qquad e^{i\left(\phi_2-\phi_3\right)} = i\epsilon_{23}.\label{prop7C}
\end{align}
Multiplying Eqs.~\eqref{prop7A} and \eqref{prop7C} gives $e^{i\left(\phi_1-\phi_3\right)}=\epsilon_{12}\epsilon_{23}$, which from Eq.~\eqref{prop7B}, must equal unity.
Thus either $\epsilon_{12}=\epsilon_{23}=1$ or $\epsilon_{12}=\epsilon_{23}=-1$.
(Note that taking a mixture of Possibilities A and B for $\epsilon_{12}$ and $\epsilon_{23}$ leads to a matrix with eigenvalues $(2/3,2/3,-1/3)$, which is not only not pure, but not positive, and so no longer a density matrix.)
If we had instead placed the imaginary factor $i$ on the $\ket{1}$ component of $\ket{\psi}$, we would have reached a conclusion of either $\epsilon_{12}=\epsilon_{13}=1$ or $\epsilon_{12}=\epsilon_{13}=-1$, thus all three $\epsilon_{12}$, $\epsilon_{13}$, and $\epsilon_{23}$ must be equal.
Clearly this argument may be trivially extended to all pairs of off-diagonals for $N>3$, through comparing chains of triples, $n\neq m\neq l$ and states of the form $\ket{\psi}=\left(\ket{n}+i\ket{m}+\ket{l}\right)/\sqrt{3}$.
\qed
\end{prf7}


As the output of Proposition \hyperlink{thm7}{7} is the input of Proposition \hyperlink{thm0}{0}, and the conclusion of the latter is the unitarity or antiunitarity of transformation $\rho\to A[\rho]$, this concludes our proof of Theorem \ref{theorem1}.

\section{Discussion}
For a finite Hilbert space, we have shown that the most general convex-structure and entropy preserving transformation on the space of density matrices is necessarily unitary or antiunitary. 
The converse statement is widely known.
If one takes the need to preserve convexity as a given, this establishes an equivalence between (anti-)unitarity and entropy conservation, so that it becomes natural to view unitary evolution, and by extension the Schr\"{o}dinger equation, as a statement of entropy conservation.

The formal statement of our result (Theorem \ref{theorem1}), sits well with other previously established symmetry theorems \cite{landsmanFoundationsQuantumTheory2017} (all of which find the dual unitary/antiunitary conclusion), the best known of which are Wigner's theorem and Kadison's theorem. 
In effect, our theorem replaces the requirement for transformation invertibility in Kadison's theorem with that of entropy conservation.
This of course makes intuitive sense, as conservation of entropy is often associated with microscopic reversibility. 

The interpretation of entropy as a scalar measure of information is widespread, and we think our result will be of most interest to researchers interested in information-forward rationalizations of quantum mechanical phenomena.
The contraposition of our theorem, not unitary implies not information conserving, means that any textbook quantum mechanical no-go theorem that arrives at its conclusion through a need to break unitarity, may instead be justified through the need to create or destroy information.
The need to create/destroy information to clone or delete quantum information has previously been shown by Horodecki \textit{et al.}~\cite{horodeckiCommonOriginNoCloning2005}. 
If, as we speculate, our theorem may be adapted to other circumstances, it could potentially be used to replace postulates of reversibility in some leading generalized probabilistic theories \cite{hardyQuantumTheoryFive2001,chiribellaQuantumTheoryNamely2012,masanesDerivationQuantumTheory2011}, and/or theories of dynamical causal structure \cite{hardyProbabilityTheoriesDynamic2005,castro-ruizDynamicsQuantumCausal2018}.

\textit{Acknowledgements---}
JRH acknowledges support from a Royal Society Research Grant (RG/R1/251590) and an EPSRC Mathematical Sciences Small Grant (UKRI3647). 
JRH and NGU acknowledge support from JRH's EPSRC Quantum Technologies Career Acceleration Fellowship (UKRI1217).

\bibliography{~/Zotero/zotero_automatically_updated.bib}

\begin{thebibliography}{29}%
\makeatletter
\providecommand \@ifxundefined [1]{%
 \@ifx{#1\undefined}
}%
\providecommand \@ifnum [1]{%
 \ifnum #1\expandafter \@firstoftwo
 \else \expandafter \@secondoftwo
 \fi
}%
\providecommand \@ifx [1]{%
 \ifx #1\expandafter \@firstoftwo
 \else \expandafter \@secondoftwo
 \fi
}%
\providecommand \natexlab [1]{#1}%
\providecommand \enquote  [1]{``#1''}%
\providecommand \bibnamefont  [1]{#1}%
\providecommand \bibfnamefont [1]{#1}%
\providecommand \citenamefont [1]{#1}%
\providecommand \href@noop [0]{\@secondoftwo}%
\providecommand \href [0]{\begingroup \@sanitize@url \@href}%
\providecommand \@href[1]{\@@startlink{#1}\@@href}%
\providecommand \@@href[1]{\endgroup#1\@@endlink}%
\providecommand \@sanitize@url [0]{\catcode `\\12\catcode `\$12\catcode
  `\&12\catcode `\#12\catcode `\^12\catcode `\_12\catcode `\%12\relax}%
\providecommand \@@startlink[1]{}%
\providecommand \@@endlink[0]{}%
\providecommand \url  [0]{\begingroup\@sanitize@url \@url }%
\providecommand \@url [1]{\endgroup\@href {#1}{\urlprefix }}%
\providecommand \urlprefix  [0]{URL }%
\providecommand \Eprint [0]{\href }%
\providecommand \doibase [0]{https://doi.org/}%
\providecommand \selectlanguage [0]{\@gobble}%
\providecommand \bibinfo  [0]{\@secondoftwo}%
\providecommand \bibfield  [0]{\@secondoftwo}%
\providecommand \translation [1]{[#1]}%
\providecommand \BibitemOpen [0]{}%
\providecommand \bibitemStop [0]{}%
\providecommand \bibitemNoStop [0]{.\EOS\space}%
\providecommand \EOS [0]{\spacefactor3000\relax}%
\providecommand \BibitemShut  [1]{\csname bibitem#1\endcsname}%
\let\auto@bib@innerbib\@empty
\bibitem [{\citenamefont
  {Landsman}(2017)}]{landsmanFoundationsQuantumTheory2017}%
  \BibitemOpen
  \bibfield  {author} {\bibinfo {author} {\bibfnamefont {K.}~\bibnamefont
  {Landsman}},\ }\href {https://doi.org/10.1007/978-3-319-51777-3} {\emph
  {\bibinfo {title} {Foundations of {{Quantum Theory}}}}},\ \bibinfo {series}
  {Fundamental {{Theories}} of {{Physics}}}, Vol.\ \bibinfo {volume} {188}\
  (\bibinfo  {publisher} {Springer International Publishing},\ \bibinfo
  {address} {Cham},\ \bibinfo {year} {2017})\BibitemShut {NoStop}%
\bibitem [{\citenamefont {Wigner}(1931)}]{wignerGruppentheorieUndIhre1931}%
  \BibitemOpen
  \bibfield  {author} {\bibinfo {author} {\bibfnamefont {E.}~\bibnamefont
  {Wigner}},\ }\href {https://doi.org/10.1007/978-3-663-02555-9} {\emph
  {\bibinfo {title} {{Gruppentheorie und ihre Anwendung auf die Quantenmechanik
  der Atomspektren}}}}\ (\bibinfo  {publisher} {Vieweg+Teubner Verlag},\
  \bibinfo {address} {Wiesbaden},\ \bibinfo {year} {1931})\BibitemShut
  {NoStop}%
\bibitem [{\citenamefont {Bargmann}(1964)}]{bargmannNoteWignersTheorem1964}%
  \BibitemOpen
  \bibfield  {author} {\bibinfo {author} {\bibfnamefont {V.}~\bibnamefont
  {Bargmann}},\ }\bibfield  {title} {\bibinfo {title} {Note on {{Wigner}}'s
  {{Theorem}} on {{Symmetry Operations}}},\ }\href
  {https://doi.org/10.1063/1.1704188} {\bibfield  {journal} {\bibinfo
  {journal} {J. Math. Phys.}\ }\textbf {\bibinfo {volume} {5}},\ \bibinfo
  {pages} {862} (\bibinfo {year} {1964})}\BibitemShut {NoStop}%
\bibitem [{\citenamefont
  {Kadison}(1965)}]{kadisonTransformationsStatesOperator1965}%
  \BibitemOpen
  \bibfield  {author} {\bibinfo {author} {\bibfnamefont {R.~V.}\ \bibnamefont
  {Kadison}},\ }\bibfield  {title} {\bibinfo {title} {Transformations of states
  in operator theory and dynamics},\ }\href
  {https://doi.org/10.1016/0040-9383(65)90075-3} {\bibfield  {journal}
  {\bibinfo  {journal} {Topology}\ }\textbf {\bibinfo {volume} {3}},\ \bibinfo
  {pages} {177} (\bibinfo {year} {1965})}\BibitemShut {NoStop}%
\bibitem [{\citenamefont
  {Hunziker}(1972)}]{hunzikerSYMMETRYOPERATIONSQUANTUM1972}%
  \BibitemOpen
  \bibfield  {author} {\bibinfo {author} {\bibfnamefont {W.}~\bibnamefont
  {Hunziker}},\ }\bibfield  {title} {\bibinfo {title} {A note on symmetry
  operations in quantum mechanics},\ }\href
  {https://doi.org/10.1007/BF02726536} {\bibfield  {journal} {\bibinfo
  {journal} {Helv. Phys. Acta}\ }\textbf {\bibinfo {volume} {45}},\ \bibinfo
  {pages} {233} (\bibinfo {year} {1972})}\BibitemShut {NoStop}%
\bibitem [{\citenamefont
  {Underwood}(2019)}]{underwoodSignaturesRelicQuantum2019}%
  \BibitemOpen
  \bibfield  {author} {\bibinfo {author} {\bibfnamefont {N.~G.}\ \bibnamefont
  {Underwood}},\ }\href {https://doi.org/10.48550/arXiv.1906.03670} {\bibinfo
  {title} {Signatures of relic quantum nonequilibrium}} (\bibinfo {year}
  {2019}),\ \Eprint {https://arxiv.org/abs/1906.03670} {arXiv:1906.03670
  [quant-ph]} \BibitemShut {NoStop}%
\bibitem [{\citenamefont
  {Underwood}(2020)}]{underwoodPrincipleInformationConservation2020}%
  \BibitemOpen
  \bibfield  {author} {\bibinfo {author} {\bibfnamefont {N.}~\bibnamefont
  {Underwood}},\ }\href {https://doi.org/10.48550/arXiv.2011.03493} {\bibinfo
  {title} {A principle of information conservation for physical laws
  ({{Hidden}} information in quantum systems?)}} (\bibinfo {year} {2020}),\
  \Eprint {https://arxiv.org/abs/2011.03493} {arXiv:2011.03493 [quant-ph]}
  \BibitemShut {NoStop}%
\bibitem [{\citenamefont
  {Shannon}(1948)}]{shannonMathematicalTheoryCommunication1949}%
  \BibitemOpen
  \bibfield  {author} {\bibinfo {author} {\bibfnamefont {C.~E.}\ \bibnamefont
  {Shannon}},\ }\bibfield  {title} {\bibinfo {title} {A mathematical theory of
  communication},\ }\href {https://doi.org/10.1002/j.1538-7305.1948.tb01338.x}
  {\bibfield  {journal} {\bibinfo  {journal} {Bell Syst. Tech. J.}\ }\textbf
  {\bibinfo {volume} {27}},\ \bibinfo {pages} {379} (\bibinfo {year}
  {1948})}\BibitemShut {NoStop}%
\bibitem [{\citenamefont
  {Stone}(1930)}]{stoneLinearTransformationsHilbert1930}%
  \BibitemOpen
  \bibfield  {author} {\bibinfo {author} {\bibfnamefont {M.~H.}\ \bibnamefont
  {Stone}},\ }\bibfield  {title} {\bibinfo {title} {Linear {{Transformations}}
  in {{Hilbert Space}}},\ }\href {https://doi.org/10.1073/pnas.16.2.172}
  {\bibfield  {journal} {\bibinfo  {journal} {Proc. Natl. Acad. Sci.}\ }\textbf
  {\bibinfo {volume} {16}},\ \bibinfo {pages} {172} (\bibinfo {year}
  {1930})}\BibitemShut {NoStop}%
\bibitem [{\citenamefont {Simon}(2015)}]{simonQuantumDynamicsAutomorphism1976}%
  \BibitemOpen
  \bibfield  {author} {\bibinfo {author} {\bibfnamefont {B.}~\bibnamefont
  {Simon}},\ }\bibfield  {title} {\bibinfo {title} {Quantum {{Dynamics}}:
  {{From Automorphism}} to {{Hamiltonian}}},\ }in\ \href
  {https://doi.org/10.1515/9781400868940-016} {\emph {\bibinfo {booktitle}
  {Studies in {{Mathematical Physics}}}}}\ (\bibinfo  {publisher} {Princeton
  University Press},\ \bibinfo {year} {2015})\BibitemShut {NoStop}%
\bibitem [{\citenamefont {Wootters}\ and\ \citenamefont
  {Zurek}(1982)}]{woottersSingleQuantumCannot1982}%
  \BibitemOpen
  \bibfield  {author} {\bibinfo {author} {\bibfnamefont {W.~K.}\ \bibnamefont
  {Wootters}}\ and\ \bibinfo {author} {\bibfnamefont {W.~H.}\ \bibnamefont
  {Zurek}},\ }\bibfield  {title} {\bibinfo {title} {A single quantum cannot be
  cloned},\ }\href {https://doi.org/10.1038/299802a0} {\bibfield  {journal}
  {\bibinfo  {journal} {Nature}\ }\textbf {\bibinfo {volume} {299}},\ \bibinfo
  {pages} {802} (\bibinfo {year} {1982})}\BibitemShut {NoStop}%
\bibitem [{\citenamefont {Dieks}(1982)}]{dieksCommunicationEPRDevices1982}%
  \BibitemOpen
  \bibfield  {author} {\bibinfo {author} {\bibfnamefont {D.}~\bibnamefont
  {Dieks}},\ }\bibfield  {title} {\bibinfo {title} {Communication by {{EPR}}
  devices},\ }\href {https://doi.org/10.1016/0375-9601(82)90084-6} {\bibfield
  {journal} {\bibinfo  {journal} {Physics Letters A}\ }\textbf {\bibinfo
  {volume} {92}},\ \bibinfo {pages} {271} (\bibinfo {year} {1982})}\BibitemShut
  {NoStop}%
\bibitem [{\citenamefont {Pati}\ and\ \citenamefont
  {Braunstein}(2000)}]{kumarpatiImpossibilityDeletingUnknown2000}%
  \BibitemOpen
  \bibfield  {author} {\bibinfo {author} {\bibfnamefont {A.~K.}\ \bibnamefont
  {Pati}}\ and\ \bibinfo {author} {\bibfnamefont {S.~L.}\ \bibnamefont
  {Braunstein}},\ }\bibfield  {title} {\bibinfo {title} {Impossibility of
  deleting an unknown quantum state},\ }\href
  {https://doi.org/10.1038/404130b0} {\bibfield  {journal} {\bibinfo  {journal}
  {Nature}\ }\textbf {\bibinfo {volume} {404}},\ \bibinfo {pages} {164}
  (\bibinfo {year} {2000})}\BibitemShut {NoStop}%
\bibitem [{\citenamefont {Horodecki}\ \emph {et~al.}(2005)\citenamefont
  {Horodecki}, \citenamefont {Horodecki}, \citenamefont {Sen(De)},\ and\
  \citenamefont {Sen}}]{horodeckiCommonOriginNoCloning2005}%
  \BibitemOpen
  \bibfield  {author} {\bibinfo {author} {\bibfnamefont {M.}~\bibnamefont
  {Horodecki}}, \bibinfo {author} {\bibfnamefont {R.}~\bibnamefont
  {Horodecki}}, \bibinfo {author} {\bibfnamefont {A.}~\bibnamefont {Sen(De)}},\
  and\ \bibinfo {author} {\bibfnamefont {U.}~\bibnamefont {Sen}},\ }\bibfield
  {title} {\bibinfo {title} {Common {{Origin}} of {{No-Cloning}} and
  {{No-Deleting Principles}} - {{Conservation}} of {{Information}}},\ }\href
  {https://doi.org/10.1007/s10701-005-8661-4} {\bibfield  {journal} {\bibinfo
  {journal} {Found Phys}\ }\textbf {\bibinfo {volume} {35}},\ \bibinfo {pages}
  {2041} (\bibinfo {year} {2005})}\BibitemShut {NoStop}%
\bibitem [{\citenamefont {Hardy}(2001)}]{hardyQuantumTheoryFive2001}%
  \BibitemOpen
  \bibfield  {author} {\bibinfo {author} {\bibfnamefont {L.}~\bibnamefont
  {Hardy}},\ }\href {https://doi.org/10.48550/arXiv.quant-ph/0101012} {\bibinfo
  {title} {Quantum {{Theory From Five Reasonable Axioms}}}} (\bibinfo {year}
  {2001}),\ \Eprint {https://arxiv.org/abs/quant-ph/0101012}
  {arXiv:quant-ph/0101012} \BibitemShut {NoStop}%
\bibitem [{\citenamefont {Chiribella}\ \emph {et~al.}(2011)\citenamefont
  {Chiribella}, \citenamefont {D'Ariano},\ and\ \citenamefont
  {Perinotti}}]{chiribellaInformationalDerivationQuantum2011}%
  \BibitemOpen
  \bibfield  {author} {\bibinfo {author} {\bibfnamefont {G.}~\bibnamefont
  {Chiribella}}, \bibinfo {author} {\bibfnamefont {G.~M.}\ \bibnamefont
  {D'Ariano}},\ and\ \bibinfo {author} {\bibfnamefont {P.}~\bibnamefont
  {Perinotti}},\ }\bibfield  {title} {\bibinfo {title} {Informational
  derivation of {{Quantum Theory}}},\ }\href
  {https://doi.org/10.1103/PhysRevA.84.012311} {\bibfield  {journal} {\bibinfo
  {journal} {Phys. Rev. A}\ }\textbf {\bibinfo {volume} {84}},\ \bibinfo
  {pages} {012311} (\bibinfo {year} {2011})},\ \Eprint
  {https://arxiv.org/abs/1011.6451} {arXiv:1011.6451 [quant-ph]} \BibitemShut
  {NoStop}%
\bibitem [{\citenamefont {Chiribella}\ \emph {et~al.}(2012)\citenamefont
  {Chiribella}, \citenamefont {D'Ariano},\ and\ \citenamefont
  {Perinotti}}]{chiribellaQuantumTheoryNamely2012}%
  \BibitemOpen
  \bibfield  {author} {\bibinfo {author} {\bibfnamefont {G.}~\bibnamefont
  {Chiribella}}, \bibinfo {author} {\bibfnamefont {G.~M.}\ \bibnamefont
  {D'Ariano}},\ and\ \bibinfo {author} {\bibfnamefont {P.}~\bibnamefont
  {Perinotti}},\ }\bibfield  {title} {\bibinfo {title} {Quantum {{Theory}},
  {{Namely}} the {{Pure}} and {{Reversible Theory}} of {{Information}}},\
  }\href {https://doi.org/10.3390/e14101877} {\bibfield  {journal} {\bibinfo
  {journal} {Entropy}\ }\textbf {\bibinfo {volume} {14}},\ \bibinfo {pages}
  {1877} (\bibinfo {year} {2012})}\BibitemShut {NoStop}%
\bibitem [{\citenamefont {Masanes}\ and\ \citenamefont
  {Mueller}(2011)}]{masanesDerivationQuantumTheory2011}%
  \BibitemOpen
  \bibfield  {author} {\bibinfo {author} {\bibfnamefont {L.}~\bibnamefont
  {Masanes}}\ and\ \bibinfo {author} {\bibfnamefont {M.~P.}\ \bibnamefont
  {Mueller}},\ }\bibfield  {title} {\bibinfo {title} {A derivation of quantum
  theory from physical requirements},\ }\href
  {https://doi.org/10.1088/1367-2630/13/6/063001} {\bibfield  {journal}
  {\bibinfo  {journal} {New J. Phys.}\ }\textbf {\bibinfo {volume} {13}},\
  \bibinfo {pages} {063001} (\bibinfo {year} {2011})},\ \Eprint
  {https://arxiv.org/abs/1004.1483} {arXiv:1004.1483 [quant-ph]} \BibitemShut
  {NoStop}%
\bibitem [{\citenamefont {Streater}\ and\ \citenamefont
  {Wightman}(1989)}]{streaterPCTSpinStatistics1989}%
  \BibitemOpen
  \bibfield  {author} {\bibinfo {author} {\bibfnamefont {R.~F.}\ \bibnamefont
  {Streater}}\ and\ \bibinfo {author} {\bibfnamefont {A.~S.}\ \bibnamefont
  {Wightman}},\ }\href {https://doi.org/10.1515/9781400884230} {\emph {\bibinfo
  {title} {{{PCT}}, {{Spin}} and {{Statistics}}, and {{All That}}}}}\ (\bibinfo
   {publisher} {Princeton University Press},\ \bibinfo {year}
  {1989})\BibitemShut {NoStop}%
\bibitem [{\citenamefont {Sudarshan}\ \emph {et~al.}(1961)\citenamefont
  {Sudarshan}, \citenamefont {Mathews},\ and\ \citenamefont
  {Rau}}]{sudarshanStochasticDynamicsQuantumMechanical1961}%
  \BibitemOpen
  \bibfield  {author} {\bibinfo {author} {\bibfnamefont {E.~C.~G.}\
  \bibnamefont {Sudarshan}}, \bibinfo {author} {\bibfnamefont {P.~M.}\
  \bibnamefont {Mathews}},\ and\ \bibinfo {author} {\bibfnamefont
  {J.}~\bibnamefont {Rau}},\ }\bibfield  {title} {\bibinfo {title} {Stochastic
  {{Dynamics}} of {{Quantum-Mechanical Systems}}},\ }\href
  {https://doi.org/10.1103/PhysRev.121.920} {\bibfield  {journal} {\bibinfo
  {journal} {Phys. Rev.}\ }\textbf {\bibinfo {volume} {121}},\ \bibinfo {pages}
  {920} (\bibinfo {year} {1961})}\BibitemShut {NoStop}%
\bibitem [{\citenamefont {Jordan}\ and\ \citenamefont
  {Sudarshan}(1961)}]{jordanDynamicalMappingsDensity1961}%
  \BibitemOpen
  \bibfield  {author} {\bibinfo {author} {\bibfnamefont {T.~F.}\ \bibnamefont
  {Jordan}}\ and\ \bibinfo {author} {\bibfnamefont {E.~C.~G.}\ \bibnamefont
  {Sudarshan}},\ }\bibfield  {title} {\bibinfo {title} {Dynamical {{Mappings}}
  of {{Density Operators}} in {{Quantum Mechanics}}},\ }\href
  {https://doi.org/10.1063/1.1724221} {\bibfield  {journal} {\bibinfo
  {journal} {J. Math. Phys.}\ }\textbf {\bibinfo {volume} {2}},\ \bibinfo
  {pages} {772} (\bibinfo {year} {1961})}\BibitemShut {NoStop}%
\bibitem [{\citenamefont {Nielsen}\ and\ \citenamefont
  {Chuang}(2010)}]{nielsenQuantumComputationQuantum2010}%
  \BibitemOpen
  \bibfield  {author} {\bibinfo {author} {\bibfnamefont {M.~A.}\ \bibnamefont
  {Nielsen}}\ and\ \bibinfo {author} {\bibfnamefont {I.~L.}\ \bibnamefont
  {Chuang}},\ }\href {https://doi.org/10.1017/CBO9780511976667} {\emph
  {\bibinfo {title} {Quantum {{Computation}} and {{Quantum Information}}: 10th
  {{Anniversary Edition}}}}}\ (\bibinfo  {publisher} {Cambridge University
  Press},\ \bibinfo {year} {2010})\BibitemShut {NoStop}%
\bibitem [{\citenamefont {Pechukas}(1994)}]{pechukasReducedDynamicsNeed1994}%
  \BibitemOpen
  \bibfield  {author} {\bibinfo {author} {\bibfnamefont {P.}~\bibnamefont
  {Pechukas}},\ }\bibfield  {title} {\bibinfo {title} {Reduced {{Dynamics Need
  Not Be Completely Positive}}},\ }\href
  {https://doi.org/10.1103/PhysRevLett.73.1060} {\bibfield  {journal} {\bibinfo
   {journal} {Phys. Rev. Lett.}\ }\textbf {\bibinfo {volume} {73}},\ \bibinfo
  {pages} {1060} (\bibinfo {year} {1994})}\BibitemShut {NoStop}%
\bibitem [{\citenamefont {Shaji}\ and\ \citenamefont
  {Sudarshan}(2005)}]{shajiWhosAfraidNot2005}%
  \BibitemOpen
  \bibfield  {author} {\bibinfo {author} {\bibfnamefont {A.}~\bibnamefont
  {Shaji}}\ and\ \bibinfo {author} {\bibfnamefont {E.}~\bibnamefont
  {Sudarshan}},\ }\bibfield  {title} {\bibinfo {title} {Who's afraid of not
  completely positive maps?},\ }\href
  {https://doi.org/10.1016/j.physleta.2005.04.029} {\bibfield  {journal}
  {\bibinfo  {journal} {Physics Letters A}\ }\textbf {\bibinfo {volume}
  {341}},\ \bibinfo {pages} {48} (\bibinfo {year} {2005})}\BibitemShut
  {NoStop}%
\bibitem [{\citenamefont {Hall}(2013)}]{hallQuantumTheoryMathematicians2013}%
  \BibitemOpen
  \bibfield  {author} {\bibinfo {author} {\bibfnamefont {B.~C.}\ \bibnamefont
  {Hall}},\ }\href {https://doi.org/10.1007/978-1-4614-7116-5} {\emph {\bibinfo
  {title} {Quantum {{Theory}} for {{Mathematicians}}}}},\ \bibinfo {series}
  {Graduate {{Texts}} in {{Mathematics}}}, Vol.\ \bibinfo {volume} {267}\
  (\bibinfo  {publisher} {Springer New York},\ \bibinfo {address} {New York,
  NY},\ \bibinfo {year} {2013})\BibitemShut {NoStop}%
\bibitem [{\citenamefont {Peres}(2002)}]{peresQuantumTheoryConcepts2002}%
  \BibitemOpen
  \bibfield  {author} {\bibinfo {author} {\bibfnamefont {A.}~\bibnamefont
  {Peres}},\ }\href {https://doi.org/10.1007/0-306-47120-5} {\emph {\bibinfo
  {title} {Quantum {{Theory}}: {{Concepts}} and {{Methods}}}}}\ (\bibinfo
  {publisher} {Kluwer},\ \bibinfo {address} {Dordrecht},\ \bibinfo {year}
  {2002})\BibitemShut {NoStop}%
\bibitem [{\citenamefont {Grabowski}\ \emph {et~al.}(2005)\citenamefont
  {Grabowski}, \citenamefont {Ku{\'s}},\ and\ \citenamefont
  {Marmo}}]{grabowskiGeometryQuantumSystems2005}%
  \BibitemOpen
  \bibfield  {author} {\bibinfo {author} {\bibfnamefont {J.}~\bibnamefont
  {Grabowski}}, \bibinfo {author} {\bibfnamefont {M.}~\bibnamefont {Ku{\'s}}},\
  and\ \bibinfo {author} {\bibfnamefont {G.}~\bibnamefont {Marmo}},\ }\bibfield
   {title} {\bibinfo {title} {Geometry of quantum systems: Density states and
  entanglement},\ }\href {https://doi.org/10.1088/0305-4470/38/47/011}
  {\bibfield  {journal} {\bibinfo  {journal} {J. Phys. A: Math. Gen.}\ }\textbf
  {\bibinfo {volume} {38}},\ \bibinfo {pages} {10217} (\bibinfo {year}
  {2005})}\BibitemShut {NoStop}%
\bibitem [{\citenamefont {Hardy}(2005)}]{hardyProbabilityTheoriesDynamic2005}%
  \BibitemOpen
  \bibfield  {author} {\bibinfo {author} {\bibfnamefont {L.}~\bibnamefont
  {Hardy}},\ }\href {https://doi.org/10.48550/arXiv.gr-qc/0509120} {\bibinfo
  {title} {Probability {{Theories}} with {{Dynamic Causal Structure}}: {{A New
  Framework}} for {{Quantum Gravity}}}} (\bibinfo {year} {2005}),\ \Eprint
  {https://arxiv.org/abs/gr-qc/0509120} {arXiv:gr-qc/0509120} \BibitemShut
  {NoStop}%
\bibitem [{\citenamefont {{Castro-Ruiz}}\ \emph {et~al.}(2018)\citenamefont
  {{Castro-Ruiz}}, \citenamefont {Giacomini},\ and\ \citenamefont
  {Brukner}}]{castro-ruizDynamicsQuantumCausal2018}%
  \BibitemOpen
  \bibfield  {author} {\bibinfo {author} {\bibfnamefont {E.}~\bibnamefont
  {{Castro-Ruiz}}}, \bibinfo {author} {\bibfnamefont {F.}~\bibnamefont
  {Giacomini}},\ and\ \bibinfo {author} {\bibfnamefont {{\v C}.}~\bibnamefont
  {Brukner}},\ }\bibfield  {title} {\bibinfo {title} {Dynamics of {{Quantum
  Causal Structures}}},\ }\href {https://doi.org/10.1103/PhysRevX.8.011047}
  {\bibfield  {journal} {\bibinfo  {journal} {Phys. Rev. X}\ }\textbf {\bibinfo
  {volume} {8}},\ \bibinfo {pages} {011047} (\bibinfo {year}
  {2018})}\BibitemShut {NoStop}%
\end{thebibliography}%
\appendix
\section{Probability conditions on Sudarshan's $A$ matrices}\label{appendix_nec_suf_for_A}
The Hermiticity of density matrix $\rho$ means that by the spectral theorem it is diagonalizable as $\rho=\sum_i p_i\ketbra{p_i}{p_i}$, where eigenvalues $p_i$ are real.
The unit trace of $\rho$ further means that $\sum_i p_i=1$, and its positive-semidefiniteness means that $p_i\geq 0$.
Hence the $p_i$ satisfy the mathematical properties of a probability density, and represent the probability of finding the system(s) described by $\rho$ in mutually orthogonal states $\ket{p_i}$.
To preserve these probability conditions, Sudarshan's $A$ matrices must preserve the Hermiticity, trace, and positive-semidefiniteness of $\rho$.
The necessary and sufficient conditions for this are for instance proved in Ref.~\cite{jordanDynamicalMappingsDensity1961}, though the notation differs from our own. 
For this reason, and to aid the reader in relating our notation to the traditional index notation introduced in \cite{sudarshanStochasticDynamicsQuantumMechanical1961}, here we summarize these proofs. 

\subsection{Conservation of Hermiticity/reality of probabilities}\label{appendix_hermiticity}
The Hermiticity conditions are
\begin{align}
&\text{Original matrix: }\label{Herm_rho}
\begin{cases}
\bra{r}\rho\ket{s}
=\left(\bra{s}\rho\ket{r}\right)^*\\
\rho_{rs}=\rho_{sr}^*
\end{cases},\\
&\text{Transformed matrix: }\label{Herm_Arho}
\begin{cases}
\bra{r}A[\rho]\ket{s}
=\left(\bra{s}A[\rho]\ket{r}\right)^*\\
\left(A[\rho]\right)_{rs}=\left(A[\rho]\right)_{sr}^*
\end{cases},\\
&\text{Condition on $A$: }\label{Herm_A}
\begin{cases}
\bra{r}A[\ketbra{n}{m}]\ket{s}=\left(\bra{s}A[\ketbra{m}{n}]\ket{r}\right)^*\\
\left(A[\ketbra{n}{m}]\right)_{rs}=\left(A[\ketbra{m}{n}]\right)_{sr}^*\\
A_{rs;nm}=A^*_{sr;mn}
\end{cases}.
\end{align}

\subsubsection{Sufficiency: \eqref{Herm_A} $\implies$ $[$\eqref{Herm_rho} $\implies$ \eqref{Herm_Arho}$]$}
An arbitrary density matrix $\rho=\sum_{nm}\rho_{nm}\ketbra{n}{m}$ transforms to 
\begin{align}
\left(A[\rho]\right)_{rs}
&=\sum_{nm}\rho_{nm}\left(A[\ketbra{n}{m}]\right)_{rs}\\
&=\sum_{nm}\rho^*_{mn}\left(A[\ketbra{n}{m}]\right)_{rs}\label{Herm_suff1}\\
&=\sum_{nm}\rho^*_{mn}\left(A[\ketbra{m}{n}]\right)_{sr}^*\label{Herm_suff2}\\
&=\sum_{nm}\rho^*_{nm}\left(A[\ketbra{n}{m}]\right)_{sr}^*
=\left(A[\rho]\right)_{sr}^*,
\end{align}
where \eqref{Herm_suff1} follows from Eq.~\eqref{Herm_rho}, and \eqref{Herm_suff2} follows from Eq.~\eqref{Herm_A}.\qed

\subsubsection{Necessity: $[$\eqref{Herm_rho} $\implies$ \eqref{Herm_Arho}$]$ $\implies$ \eqref{Herm_A}}
First consider pure state $\rho=\ketbra{n}{n}$. This satisfies Eq.~\eqref{Herm_rho}. The corresponding condition on the transformed matrix [Eq.~\eqref{Herm_Arho}] is $\left(A[\ketbra{n}{n}]\right)_{rs}=\left(A[\ketbra{n}{n}]\right)_{sr}^*$.
Second consider the pure state corresponding to quantum state $\frac{1}{\sqrt{2}}\left(\ket{n}+\ket{m}\right)$ with ($n\neq m$), which transforms to
\begin{align}
\frac{1}{2}A\left[\ketbra{n}{n}+\ketbra{m}{m}\right]+ \frac{1}{2}A\left[\ketbra{n}{m}+\ketbra{m}{n}\right].
\end{align}
As by condition \eqref{Herm_Arho} this is overall Hermitian, and we have just established that the first term is Hermitian, the second term is also required to be Hermitian,
\begin{align}\label{Herm_nec1}
\left(A[\ketbra{n}{m}]\right)_{rs}+\left(A[\ketbra{m}{n}]\right)_{rs}
=\left(A[\ketbra{n}{m}]\right)_{sr}^*+\left(A[\ketbra{m}{n}]\right)_{sr}^*.
\end{align}
Third, repeating this argument applied to quantum state $\frac{1}{\sqrt{2}}\left(\ket{n}+i\ket{m}\right)$ with ($n\neq m$), results in
\begin{align}\label{Herm_nec2}
&-i\left(A[\ketbra{n}{m}]\right)_{rs}+i\left(A[\ketbra{m}{n}]\right)_{rs}\nonumber\\
=&i\left(A[\ketbra{n}{m}]\right)_{sr}^*-i\left(A[\ketbra{m}{n}]\right)_{sr}^*.
\end{align}
Finally, a combination of equations \eqref{Herm_nec1} and \eqref{Herm_nec2} results in 
$\left(A[\ketbra{n}{m}]\right)_{rs}=\left(A[\ketbra{m}{n}]\right)_{sr}^*$ for $n\neq m$.\qed

\subsection{Conservation of normalisation of probabilities/trace}\label{appendix_normalisation}
The unit trace conditions are
\begin{align}
&\text{Original matrix: }\label{Trace_rho}
\begin{cases}
1=\tr(\rho)\\
=\sum_r\bra{r}\rho\ket{r}\\
=\sum_r\rho_{rr}
\end{cases},\\
&\text{Transformed matrix: }\label{Trace_Arho}
\begin{cases}
1=\tr(A[\rho])\\
=\sum_r\bra{r}A[\rho]\ket{r}\\
=\sum_r\left(A[\rho]\right)_{rr}
\end{cases},\\
&\text{Condition on A: }\label{Trace_A}
\begin{cases}
\delta_{nm}=\tr\left(A[\ketbra{n}{m}]\right)\\
=\sum_r\bra{r}A[\ketbra{n}{m}]\ket{r}\\
=\sum_r\left(A[\ketbra{n}{m}]\right)_{rr}\\
=\sum_rA_{rr;nm}
\end{cases}.
\end{align}

\subsubsection{Sufficiency: \eqref{Trace_A} $\implies$ $[$\eqref{Trace_rho} $\implies$ \eqref{Trace_Arho}$]$}
The trace of the transformation of an arbitrary matrix $\rho=\sum_{nm}\rho_{nm}\ketbra{n}{m}$ is
\begin{align}
\tr(A[\rho])
=\sum_r\left(A[\rho]\right)_{rr}
&=\sum_r\sum_{nm}\rho_{nm}\left(A[\ketbra{n}{m}]\right)_{rr}\\
&=\sum_{nm}\rho_{nm}\delta_{nm}\label{Trace_suff1}
=\tr\rho,
\end{align}
where equality \eqref{Trace_suff1} is reached using property \eqref{Trace_A}.\qed

\subsubsection{Necessity: $[$\eqref{Trace_rho} $\implies$ \eqref{Trace_Arho}$]$ $\implies$ \eqref{Trace_A}}
First consider pure state $\rho=\ketbra{n}{n}$. This satisfies Eq.~\eqref{Trace_rho}. The corresponding condition on the transformed matrix [Eq.~\eqref{Trace_Arho}] is $\tr\left(A[\ketbra{n}{n}]\right)=1$.

Second consider the pure state corresponding to quantum state $\frac{1}{\sqrt{2}}\left(\ket{n}+\ket{m}\right)$ with ($n\neq m$). The corresponding Eq.~\eqref{Trace_Arho} is
\begin{align}
1=\frac{1}{2}\tr\left(A[\ketbra{n}{n}]\right)
+\frac{1}{2}\tr\left(A[\ketbra{m}{m}]\right)\\
+\frac{1}{2}\tr\left(A[\ketbra{n}{m}]\right)
+\frac{1}{2}\tr\left(A[\ketbra{m}{n}]\right),
\end{align}
where the first two terms on the RHS sum to unity so that 
\begin{align}\label{Trace_nec1}
0=\tr\left(A[\ketbra{n}{m}]\right)
+\tr\left(A[\ketbra{m}{n}]\right).
\end{align}
Third, repeating this argument applied to quantum state $\frac{1}{\sqrt{2}}\left(\ket{n}+i\ket{m}\right)$ with ($n\neq m$), results in
\begin{align}\label{Trace_nec2}
0=-\tr\left(A[\ketbra{n}{m}]\right)+\tr\left(A[\ketbra{m}{n}]\right)
\end{align}
Finally, comparison of Eqs.~\eqref{Trace_nec1} and \eqref{Trace_nec2} results in $\tr\left(A[\ketbra{n}{m}]\right)=0$ for $n\neq m$. \qed

\subsection{Conservation of non-negative probabilities/positive semi-definiteness}\label{appendix_PSD}
For all $\ket{\psi}$ and $\ket{\phi}$ (not necessarily normalised),
\begin{align}
&\text{Original matrix:}\quad\label{PSD_rho}
\bra{\psi}\rho\ket{\psi}\geq 0,\\
&\text{Transformed matrix:}\quad\label{PSD_Arho}
\bra{\psi}A[\rho]\ket{\psi} \geq 0, \\
&\text{Condition on A:}\quad\label{PSD_A}
\bra{\psi}A[\ketbra{\phi}{\phi}]\ket{\psi} \geq 0 .
\end{align}

\subsubsection{Sufficiency: \eqref{PSD_A} $\implies$ $[$\eqref{PSD_rho} $\implies$ \eqref{PSD_Arho}$]$}
Any density matrix is expressible as a (not necessarily unique) convex combination of pure states, $\rho=\sum_i p_i\ketbra{\phi_i}{\phi_i}$, $p_i>0$.
So,
\begin{align}
\bra{\psi}A[\rho]\ket{\psi}
&=\bra{\psi}A\Big[\sum_i p_i\ketbra{\phi_i}{\phi_i}\Big]\ket{\psi}\\
&=\sum_i p_i\bra{\psi}A\left[\ketbra{\phi_i}{\phi_i}\right]\ket{\psi}\geq0
\end{align}
where to find the final inequality we have used Eq.~\eqref{PSD_A}.\qed

\subsubsection{Necessity: $[$\eqref{PSD_rho} $\implies$ \eqref{PSD_Arho}$]$ $\implies$ \eqref{PSD_A}}
The statement $[$\eqref{PSD_rho} $\implies$ \eqref{PSD_Arho}$]$ means that for any density matrix $\rho$, the product $\bra{\psi}A[\rho]\ket{\psi}$ is non-negative. 
This includes pure states $\rho=\ketbra{\phi}{\phi}$, so $\bra{\psi}A[\ketbra{\phi}{\phi}]\ket{\psi}\geq0$, and since any complex factor may be attached to $\ket{\psi}$ or $\ket{\phi}$ without changing the sign of the product, and these span Hilbert space $H$, this applies for all $\ket{\psi}$ and $\ket{\phi}$. \qed

\end{document}